\documentclass[apj]{emulateapj}
\usepackage{apjfonts}
\shorttitle{IRAC OBSERVATIONS OF LOW-MASS WDs}
\shortauthors{KILIC, BROWN, \& McLEOD}

\begin{document}

\title{A {\em Spitzer} Search for Substellar Companions to Low Mass White Dwarfs\footnote{Based partly on observations obtained at the MMT Observatory, a joint facility of the Smithsonian Institution and the University of Arizona.}}

\author{Mukremin Kilic\altaffilmark{1,2}, Warren R. Brown, and B. McLeod}

\affil{Smithsonian Astrophysical Observatory, 60 Garden St., Cambridge, MA 02138}
\altaffiltext{1}{\em Spitzer Fellow}
\altaffiltext{2}{mkilic@cfa.harvard.edu}

\begin{abstract}

The formation scenarios for {\it single} low-mass ($M<0.45M_{\odot}$) white dwarfs include enhanced mass loss from a metal-rich
progenitor star or a common envelope phase of a solar-like star with a close-in massive planet
or a brown dwarf. Both scenarios suggest that low-mass white dwarfs may have planets.
Here, we present a {\em Spitzer} IRAC search for substellar and planetary mass companions to 14 low-mass white dwarfs.
One of our targets, HS 1653+7753, displays near- and mid-infrared flux excess. However, follow-up MMT observations show that
this excess is due to a nearby resolved source, which is mostly likely a background object.
Another target, PG 2257+162, shows flux excess compatible with a late-type stellar companion. 
We do not detect substellar companions to any of the remaining targets.
In addition, eight of these stars do not show any radial velocity variations, ruling out stellar mass companions including other white
dwarfs.
We conclude that a significant fraction of the low-mass white dwarfs in our sample do not have stellar or massive brown dwarf companions.

\end{abstract}


\section{INTRODUCTION}

The fate of planetary systems around Sun-like stars can be studied by
finding planets around their white dwarf (WD) remnants.
WDs are physically small ($R\sim0.01R_\odot$),
which sharply limits their luminosity in the infrared, where the
planet spectrum peaks \citep{burleigh02,ignace01}. Compared to main-sequence (MS) stars, the significant gain in
contrast makes WDs excellent targets for photometric searches for planetary companions.
However, to this day, no planet detection around WDs has been reported.
{\em Spitzer} IRAC imaging surveys by \citet{mullally07}, \citet{debes07}, \citet{farihi08}, and \citet{kilic09b},
and near-infrared imaging surveys by \citet{friedrich06}, \citet{burleigh08}, and \citet{hogan09} were unsuccessful
in finding planetary mass companions to WDs with limits typically between 5 and $10 M_J$. 
Here we focus on using low-mass WDs to provide additional leverage
for detecting planetary mass companions.

\subsection{CURRENT CONSTRAINTS}

Timing measurements of pulsating WDs is an alternative, powerful method to search for planets \citep{winget03}, but
this method can be used only for a limited sample of stars.
There is a promising signature of a $M\sin i= 2.4 M_J$ planet in the pulsation timing measurements of the WD GD 66 \citep{mullally09}.
However, a full orbit has not been observed, and this detection remains provisional (F. Mullally 2009, priv. comm.).

The discovery of circumstellar debris disks around more than a dozen WDs imply that
1\% to 3\% of WDs with cooling ages $<$ 500 Myr have disks \citep[see][and references therein]{farihi09}.
These disks most likely formed by tidally disrupted asteroids \citep{jura03}. Hence, there is indirect
evidence that at least a few percent of WDs have remnant planetary systems. Of course, the observed debris
disk fraction may represent only the planetary systems that become unstable during post-MS evolution, 
and the true fraction of WDs with planets may be much higher.

The reason for the lack of known planetary companions to WDs may be a target selection effect \citep{kilic07b}.
\citet{fischer05} find that less than 3\% of the stars with $-$0.5 $<$ [Fe/H] $<$ 0.0 have
Doppler-detected planets, whereas 25\% of the stars with [Fe/H] $>$ +0.3 have gas giant planets.
Based on the age-metallicity relation derived by \citet{reid07}, \citet{kilic07b} find that only 3\% of the stars in
the solar neighborhood have had [Fe/H] $>$ +0.3 in the past 10 Gyr. Therefore, a random selection of WDs would be dominated
by WDs that come from low metallicity progenitors. The best way to find planetary companions to WDs may be
to search for companions to WDs that form from metal-rich stars.

\subsection{LOW-MASS WHITE DWARFS}

Low-mass helium-core WDs ($M < 0.45\ M_\odot$) can be produced from interacting binary systems, and traditionally all of them have been attributed
to this channel. However, a low-mass WD can also result from a single star that experiences severe mass loss on the first ascent
giant branch. \citet{kalirai07} identified several low-mass WDs in the upper part of the WD cooling sequence of the
metal-rich cluster NGC 6791, and they argued
that there is a mechanism in clusters to produce low-mass WDs without
requiring binary star interactions.
However, this argument is now challenged by \citet{bedin08}, who argue that the bright peak observed in the WD luminosity fuction
of this cluster is better explained as binary WDs. \citet{bedin08} also argue that the low-mass WDs found by \citet{kalirai07}
can be explained as the product of binary star evolution.

Unlike the relatively faint low-mass WDs in NGC 6791, it is possible to obtain radial velocity measurements of low-mass WDs
in the field and test the binary formation scenario. 
Around half the low-mass WDs observed by \citet{maxted00} lack detectable radial velocity companions.
The ESO SNIa Progenitor Survey (SPY) searched for radial velocity variations in more than a thousand WDs
using the Very Large Telescope \citep{napiwotzki01}. They found that about 58\%
of the low-mass WDs in their sample do not show radial velocity variations \citep{napiwotzki07}.
Their detailed analysis of the detection probability of MS, WD, and brown dwarf companions show
that their observations of low-mass WDs are inconsistent with the idea that all of them reside in close binary systems.
The SPY observations are less sensitive to brown dwarf companions, because of their smaller mass, but even in this
case 100\% binary fraction rate can be ruled out (R. Napiwotzki 2007, private communication).
\citet{kilic07b} argue that these single low-mass WDs
might come from old metal-rich stars that truncate their evolution prior to the helium flash from severe mass loss.
The fraction of single low-mass WDs among the field WD sample is also compatible with the fraction of metal-rich dwarfs
among the field dwarf population.

A strong implication of the \citet{kilic07b} scenario is that single low-mass WDs are good targets for
planet searches because they may arise from metal-rich progenitors.
This leads to an inversion of prior strategies for seeking planetary companions to WDs, which avoid low-mass WDs on the grounds that they are
the products of binary interactions.
Instead, some of these objects may be the descendants of metal-rich stars that are known to have a high frequency of
planet detection. If the progenitors of these single low-mass WDs have planets, the planets are more likely
to survive the post-MS evolution, as these He-core WDs do not go through the asymptotic giant branch phase.

An alternative formation scenario for single low-mass WDs is proposed by \citet{nelemans98}. 
In this scenario, low-mass WDs form from Sun-like stars with close planetary companions.
A common envelope phase with a massive planet can expel the stellar envelope and form a low-mass WD.
The planet may (not) survive the common envelope phase. However, if there are additional planets in wide orbits,
they are likely to survive \citep{livio83,burleigh02,duncan98}.

Both scenarios proposed by \citet{nelemans98} and \citet{kilic07b} suggest that
low-mass WDs may have planets. Here we present a {\em Spitzer}
IRAC search for substellar companions to 14 low-mass WDs.
Our sample selection and observations are discussed in Section 2, whereas the detectability of planets around
our targets and the results of our search are discussed in Section 3 and 4.

\section{OBSERVATIONS}

\citet{maxted00} presented a radial velocity survey of 71 WDs including 13 low-mass WDs
with no detectable radial velocity variations. 
\citet{hansen06} and \citet{mullally07} obtained IRAC observations of 3 of these stars (WD 0316+345, WD 1353+409, and WD 1614+136).
We target the remaining 10 single low-mass WDs for follow-up observations.
In addition, we target 4 low-mass WDs from the 
Palomar-Green Survey \citep{liebert05}.
\citet{farihi05} presented near-infrared photometry for two of these stars
which did not reveal any companions. 
The physical parameters of our targets, including masses, distances, and WD cooling ages,
are presented in Table 1. All of our targets are hydrogen-rich DA WDs.

%TABLE1
\begin{deluxetable}{lccrrr}
\tabletypesize{\footnotesize}
\tablecolumns{6}
\tablewidth{0pt}
\tablecaption{Low Mass WD Targets}
\tablehead{
\colhead{Object}&
\colhead{$T_{\rm eff}$}&
\colhead{$M_{\rm WD}$}&
\colhead{$d$}&
\colhead{$\tau_{\rm WD}$}&
\colhead{Reference}
\\
        & (K) & ($M_\odot$) & (pc) & (Myr) & }
\startdata
PG 0132+254 & 19960 & 0.41 & 180 & 60 & 1 \\
PG 0237+242 & 69160 & 0.40 & 881 & 20 & 1 \\
WD 0339+523 & 12640 & 0.33 & 103 & 180 & 2 \\
WD 0437+152 & 10410 & 0.38 & 153 & 340 & 3 \\
WD 0453+418 & 13660 & 0.44 & 44  & 150 & 2 \\
WD 0509$-$007 & 32200 & 0.40 & 117 & 10 & 4 \\
PG 0808+595 & 27330 & 0.42 & 238 & 10 & 1 \\
PG 0943+441 & 12820 & 0.41 & 30 & 340 & 1 \\
WD 0950$-$572 & 12400 & 0.44 & 58 & 230 & 2 \\
PG 1229$-$013 & 19430 & 0.41 & 59 & 70 & 1 \\
HS 1653+7753 & 29400 & 0.32 & 180 & 10 & 5 \\
PG 1654+637 & 15070 & 0.44 & 98 & 200 & 1 \\
PG 2226+061 & 15280 & 0.44 & 65 & 190 & 1 \\
PG 2257+162 & 24580 & 0.43 & 212 & 20 & 1 \\
\enddata
\tablecomments{1- \citet{liebert05}, 2- Gianninas et al. (2005), 3- Moran (1999),
4- Vennes et al. (1997), 5- Homeier et al. (1998)}
\end{deluxetable}

Observations reported here were obtained as part of the Cycle 5 Spitzer Fellowship Program 474.
We obtained 3.6, 4.5, 5.8, and 8 $\mu$m images with integration times of 30 or 100 seconds per dither,
with five or nine dithers per object. We use the IRAF PHOT and IDL astrolib packages to perform
aperture photometry on the individual BCD frames from the latest available IRAC reduction pipeline (17.2 or 18.5 for our targets).
Since our targets are relatively faint, we use the smallest aperture (2 pixels) for which there are published
aperture corrections. Following the IRAC calibration procedure, corrections for the location of the source in the array
were taken into account before averaging the fluxes of each of the dithered frames at each wavelength.
Channel 1 (3.6$\mu$m) photometry is also corrected for the pixel-phase-dependence.
The results from IRAF and IDL reductions are consistent within the errors.
The photometric error bars are estimated from the observed scatter in the 5 (or 9) images
corresponding to the dither positions. We also add the 3\% absolute calibration error in quadrature.
Finally, we divide the estimated fluxes by the color corrections for a Rayleigh-Jeans spectrum \citep{reach05}.
These corrections are 1.0111, 1.0121, 1.0155, and 1.0337 for the 3.6, 4.5, 5.8,
and 8 $\mu$m bands, respectively.

We present the IRAC photometry of our targets in Table 2. Most
of our targets are detected in 2MASS, at least in the
$J$ band. Several targets also have Sloan Digital Sky Survey (SDSS) photometry available.
We obtained $V$ and $I$ band photometry
of HS 1653+7753 using the MDM 2.4m telescope equipped with the Echelle CCD
on UT 2009 April 5. We use observations of the standard star field PG 1633+099 \citep{landolt92}
to calibrate the photometry. The photometric reductions were performed by J. Thorstensen, and kindly made available to us.
HS 1653+7753 has $V=$ 15.08 mag and $I=$ 15.33 mag. We estimate an internal accuracy of 0.01 mag for the photometry,
but to be conservative, we adopt errors of 0.03 mag.
H. C. Harris kindly provided us unpublished $BVI$ photometry of WD 0339+523. We use the optical photometry from
\citet{mccook06} for the remaining targets.

%TABLE2
\begin{deluxetable}{lrrrr}
\tabletypesize{\scriptsize}
\tablecolumns{5}
\tablewidth{0pt} 
\tablecaption{IRAC Photometry of Low Mass WDs}
\tablehead{
\colhead{Object}&
\colhead{3.6$\mu$m}&
\colhead{4.5$\mu$m}& 
\colhead{5.8$\mu$m}&
\colhead{8.0$\mu$m}
\\
        & ($\mu$Jy) & ($\mu$Jy) & ($\mu$Jy) & ($\mu$Jy) }
\startdata
PG 0132+254 & 54.7 $\pm$ 3.2 & 32.9 $\pm$ 1.4 & 21.1 $\pm$ 11.0 & 25.6 $\pm$ 11.2 \\
PG 0237+242 & 54.2 $\pm$ 3.4 & 34.4 $\pm$ 1.8 & 24.2 $\pm$ 10.8 & 22.5 $\pm$ 8.4  \\
WD 0339+523 & 103.9 $\pm$ 5.0 & 66.1 $\pm$ 3.8 & 33.7 $\pm$ 19.2 & 23.7 $\pm$ 20.6 \\
WD 0437+152 & 79.0 $\pm$ 4.1 & 49.2 $\pm$ 2.2 & 31.3 $\pm$ 17.7 & 22.0 $\pm$ 19.3 \\
WD 0453+418 & 449.7 $\pm$ 14.4 & 287.3 $\pm$ 10.4 & 176.7 $\pm$ 25.9 & 102.9 $\pm$ 46.8 \\
WD 0509$-$007 & 333.3 $\pm$ 11.1 & 213.4 $\pm$ 11.5 & 141.8 $\pm$ 18.9 & 73.5 $\pm$ 18.3 \\
PG 0808+595 & 48.3 $\pm$ 5.3 & 27.8 $\pm$ 2.4 & 13.5 $\pm$ 13.2 & 13.4 $\pm$ 4.2 \\
PG 0943+441 & 881.2 $\pm$ 32.4 & 549.1 $\pm$ 19.4 & 367.1 $\pm$ 16.7 & 212.7 $\pm$ 19.7 \\
WD 0950$-$572 & 269.6 $\pm$ 14.4 & 173.7 $\pm$ 7.7 & 121.9 $\pm$ 19.3 & 121.9 $\pm$ 40.1 \\ 
PG 1229$-$013 & 237.6 $\pm$ 9.2 & 150.3 $\pm$ 6.8 & 80.4 $\pm$ 9.4 & 73.8 $\pm$ 32.1 \\
HS 1653+7753 & 161.3 $\pm$ 5.1 & 106.5 $\pm$ 3.9 & 66.9 $\pm$ 5.7 & 39.5 $\pm$ 7.9 \\
PG 1654+637 & 89.6 $\pm$ 4.7 & 53.9 $\pm$ 2.3 & 37.1 $\pm$ 14.6 & 15.2 $\pm$ 11.1 \\
PG 2226+061 & 198.6 $\pm$ 14.6 & 122.1 $\pm$ 7.1 & 74.5 $\pm$ 24.6 & 41.2 $\pm$ 21.0 \\
PG 2257+162 & 410.7 $\pm$ 17.9 & 282.4 $\pm$ 14.1 & 190.1 $\pm$ 30.4 & 113.2 $\pm$ 36.6 \\
\enddata
\end{deluxetable} 

In addition, we obtained $JHK$ imaging observations of HS 1653+7753 using the MMT and Magellan Infrared Spectrograph \citep[MMIRS;][]{mcleod04}
on the MMT on UT 2009 June 13.
The observations were obtained under cloudy conditions, and they included 10 exposures of 30 s in $J$, 10 exposures of 20 s in $H$,
and 25 exposures of 45 s in $K$. 

\section{RESULTS}

Figure 1 presents the spectral energy distributions of twelve
of our fourteen low-mass WDs. Two WDs, with more complicated SEDs,
are presented below.
D. Koester kindly provided pure-hydrogen atmosphere WD model spectra for our targets.
We perform synthetic photometry on these models using the appropriate transmission
curves for the $BVRI$, 
SDSS, 2MASS, and IRAC filters. The resulting fluxes are then compared with the observations to find
the normalization factor for the WD models. We weight fluxes by their associated error bars.
The solid lines in Figure 1 present the appropriate WD model for each star normalized to match the observations.
The use of optical and near-infrared photometry helps us constrain the predicted mid-infrared photospheric fluxes
for WDs. The mid-infrared
photometry of all but one of the twelve WDs presented in this figure is consistent with the predicted
photospheric flux from WDs; we do not detect excess mid-infrared flux from planets or substellar companions.
The thermal emission from planetary companions would be detected most strongly at 4.5 $\mu$m \citep{burrows03}.

\begin{figure}
\includegraphics[angle=-90,scale=.37]{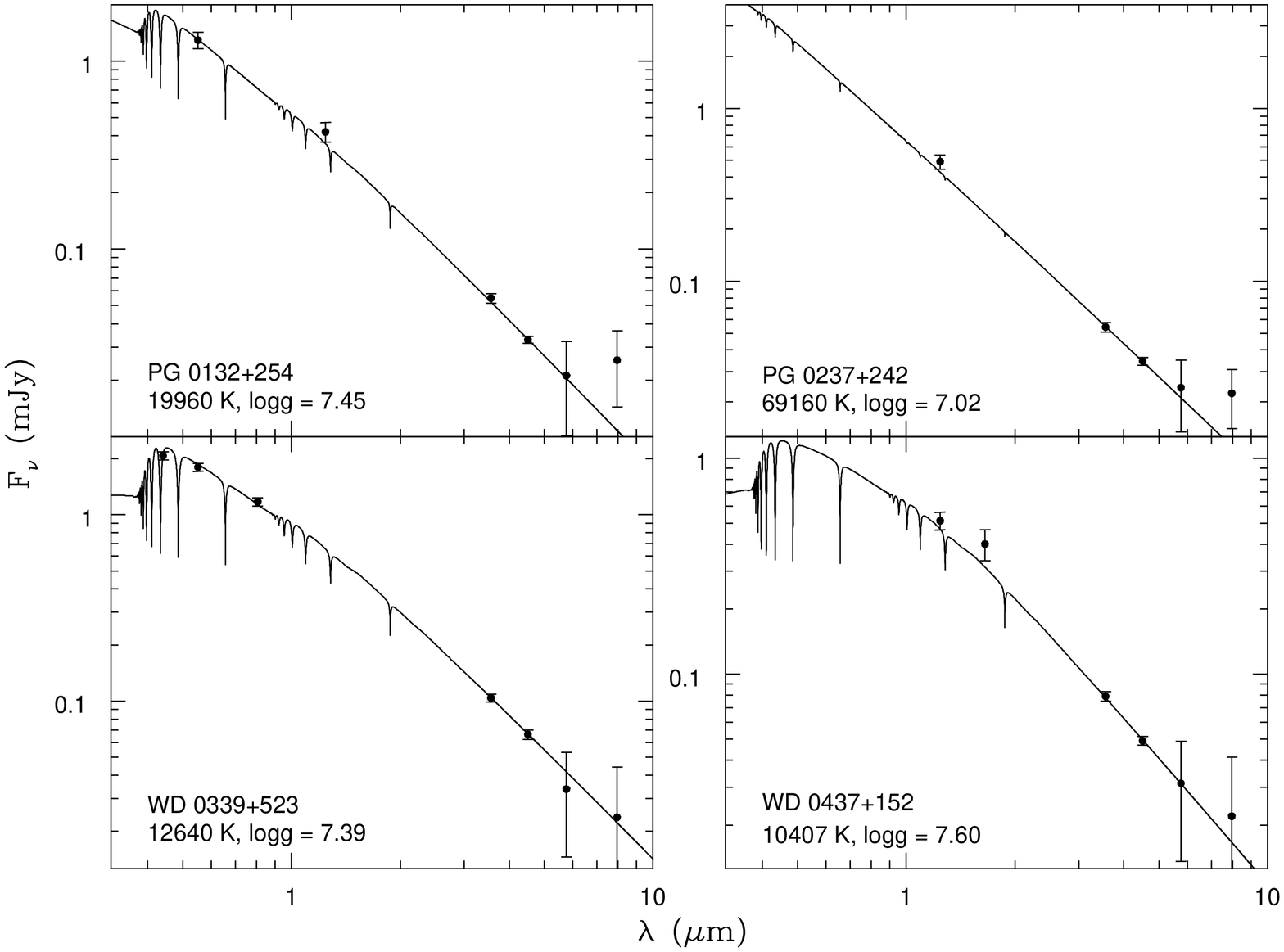}
\includegraphics[angle=-90,scale=.37]{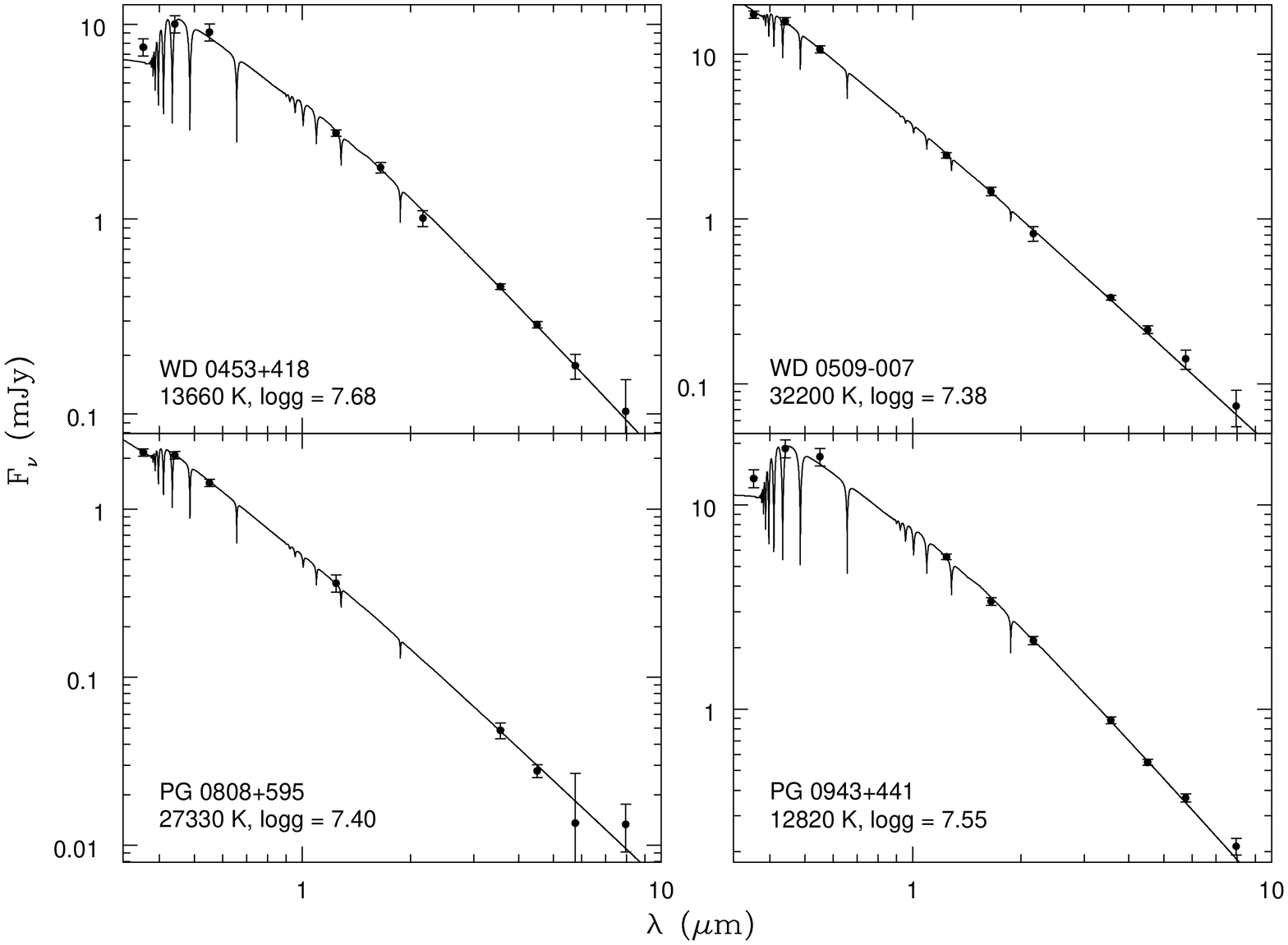}
\includegraphics[angle=-90,scale=.37]{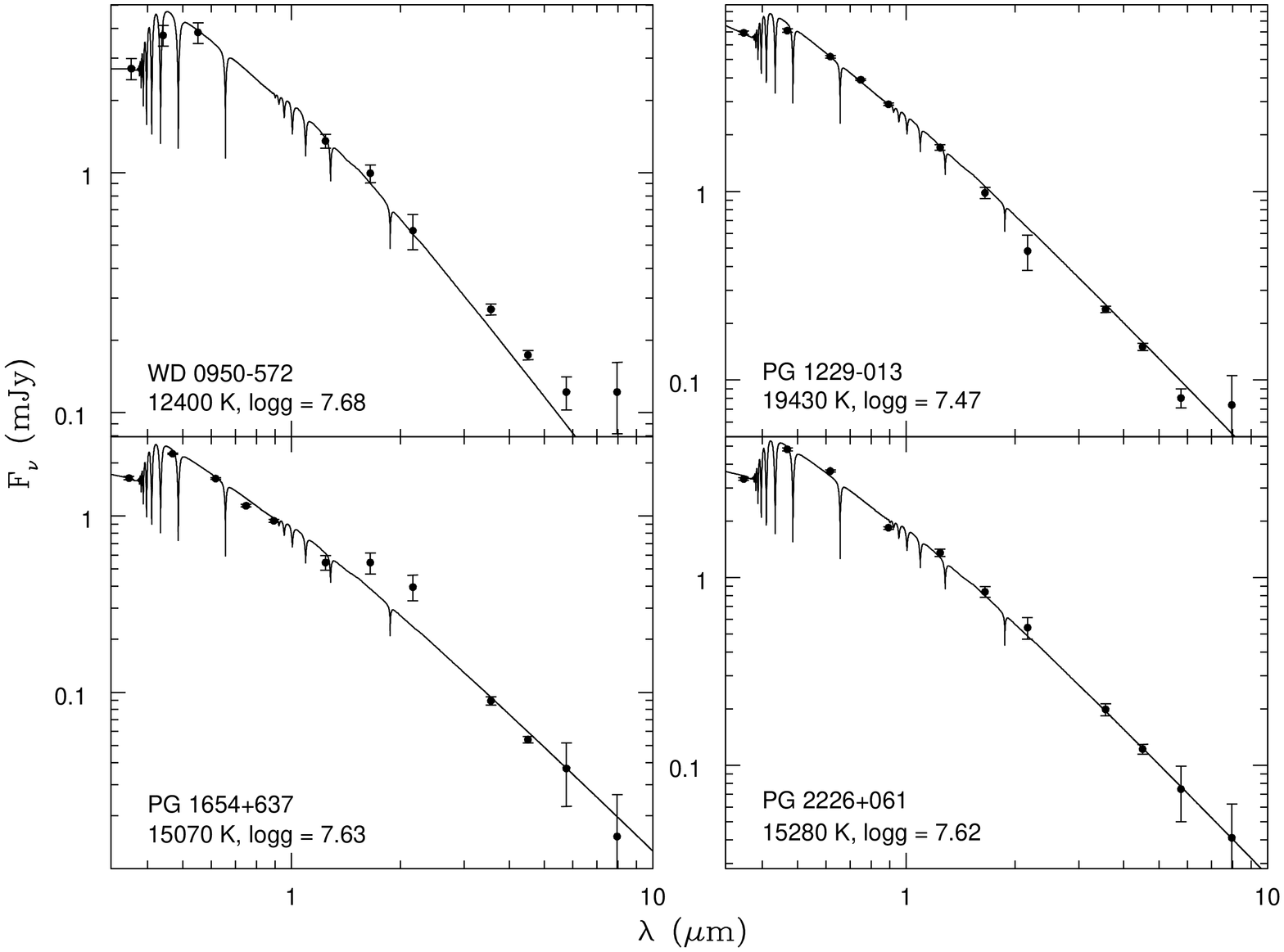}
\caption{Spectral energy distributions of 12 low-mass WDs. The expected photospheric flux
from the WDs are shown as solid lines (D. Koester 2009, priv. comm.).}
\end{figure}

One of the objects in Figure 1, WD 0950$-$572 shows a mid-infrared flux excess. Figure 2 
presents a contour map of the immediate field around this WD; it is near a bright star and even the smallest
aperture that we use (2 pixels) is contaminated by flux from this nearby bright star. Therefore, the observed excess flux
is most likely not intrinsic to the WD environment.

\begin{figure}
\plotone{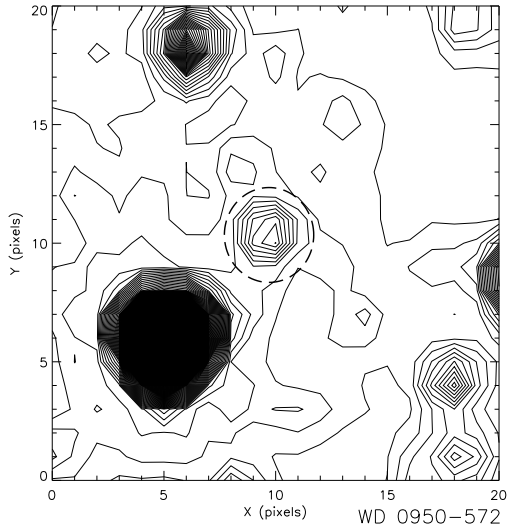}
\caption{Contour map of the immediate field around WD 0950$-$572. The map covers 20 x 20 pixels, which
corresponds to 24" x 24". Dashed circle marks a 2 pixel aperture centered on the WD.}
\end{figure}

\begin{figure}[b]
\hspace{-0.5in}
\includegraphics[angle=-90,scale=.40]{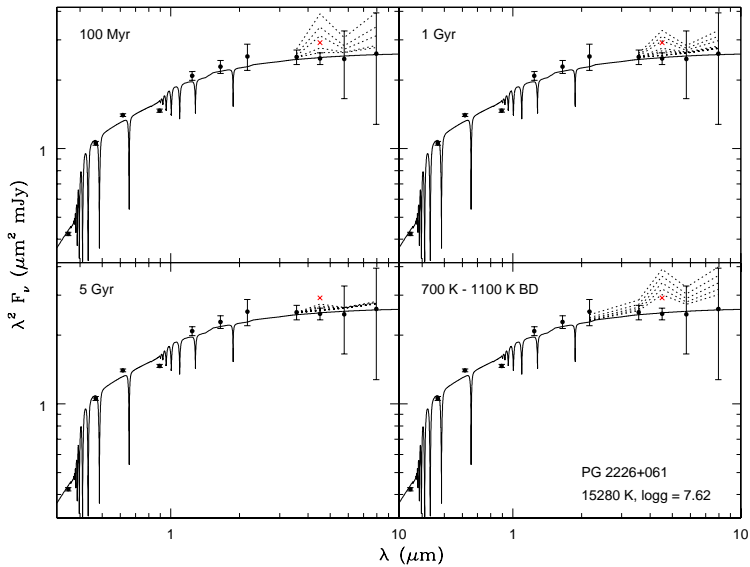}
\caption{Spectral energy distribution of PG 2226+061 compared to the expected photospheric flux from the WD (solid line),
and from planetary ($1-25M_J$) and brown dwarf companions \citep[dotted lines,][]{burrows03,burrows06}.
The red cross marks the 3$\sigma$ upper limit of the 4.5 $\mu$m photometry.}
\end{figure}

\subsection{DETECTABILITY OF PLANETS}

The usual method for estimating the upper mass limit for substellar
companions in photometric surveys is to calculate the total age of the
white dwarf, i.e. the WD cooling age plus the MS lifetime, and then
use evolutionary models for substellar objects and giant planets to
predict their brightness at the age and distance of the WD
\citep[see for example][]{hogan09,farihi08}.
But there is a problem for low mass WDs: if they form through
common envelope evolution, it is not possible to know their
progenitor mass and their MS lifetimes.
Hence only a lower limit to the age of the system, i.e. the WD cooling age, is known.
Even if the low-mass WDs form directly from single
metal-rich progenitors, currently we do not have initial-final mass relations
that would give us progenitor masses and lifetimes for such systems.
Therefore, all we can do is to calculate companion upper mass limits
for a range of possible ages starting from the WD cooling ages to several Gyr.
All of the objects in our sample are hotter than 10,000 K (Table 1), therefore their WD cooling ages are less than 350 Myr \citep{bergeron95}.

A relatively massive planet or brown dwarf can stay hot and easily detectable by $Spitzer$ for billions of years.
Some brown dwarfs or planets may also be reheated during the red giant phase of the progenitor's evolution, either by direct
radiation or by impacts from planets and debris belts destabilized during stellar mass loss \citep{parriott98,debes02}.
Collisions between two planets may create mergers that would essentially reset the cooling clock of the planet and make
the remnant planet overluminous by a factor of a hundred. However,
\citet{debes02} predict that only about 2\% of young WDs may have overluminous planetary companions that form through collisions.

\citet{melis09} report the discovery of a unique first-ascent giant star with a substantial dust disk. They suggest that the circumstellar
material around this star may originate from a common-envelope phase with a low-mass stellar or substellar companion.
The gaseous
and dusty material is similar to the primordial disks observed around T Tauri stars. They speculate that these disks may form a second
generation of planets around first-ascent giant stars. Single low-mass WDs may descend from such stars \citep{nelemans98}.
The discovery of a close-in substellar companion to the low-mass
WD 0137$-$049 implies that this substellar companion has gone through a common-envelope phase that resulted in the ejection
of the envelope of the progenitor first-ascent giant star \citep{maxted06}.
If the first-ascent giant stars with gaseous/dusty disks are common and if these stars form low-mass single WDs, then these WDs
may have second generation planets as well. These planets would be as young as the cooling age of the WD, and therefore
warmer and more luminous. 

\subsection{LIMITS ON COMPANIONS}

For the 11 stars presented in Figure 1 (excluding WD 0950$-$572),
we use \citet{baraffe03} and \citet{burrows03,burrows06}
models to put upper limits on companions that would escape detection.
I. Baraffe kindly provided us substellar cooling models updated to include
fluxes in the IRAC bandpasses for 0.1-5 Gyr old objects with $M=$ 1-100 $M_J$.

Figure 3 shows an example of predicted excess flux from planetary and brown dwarf companions around one of our targets,
PG 2226+061. By demanding a 3$\sigma$ excess from any possible companion, we exclude 100 Myr old,
$M > 4 M_J$ planets around PG 2226+061. Depending on the unknown ages of the possible planetary companions,
this limit goes up to more than 30 $M_J$ for 5 Gyr or older planets.
In all cases, we are able to rule out brown dwarf companions with $T_{\rm eff}\geq 700$ K around this WD.

We achieve similar limits on substellar companions for the rest of our sample. These limits and our search radii
are presented in Table 3. Since the total ages (WD cooling age + MS lifetime) are unknown, we present
companion mass limits for a range of ages.
If the planets are as old as the WD cooling ages (20-340 Myr), we can rule out planets more massive than $5M_J$ to $10M_J$
around 10 of our targets; second-generation massive planets are not common around low-mass WDs.
However, if the planets are as old as 5 Gyr, only massive brown dwarf companions are ruled out.

%TABLE3
\begin{deluxetable}{lrllllr}
\tabletypesize{\scriptsize}
\tablecolumns{7}
\tablewidth{0pt}
\tablecaption{Ruled out Companions}
\tablehead{
\colhead{Object}&
\colhead{R}&
\colhead{0.1 Gyr}&
\colhead{0.5 Gyr}&
\colhead{1.0 Gyr}&
\colhead{5.0 Gyr}&
\colhead{BD}
\\
  &  (AU)  & ($M_J$) & ($M_J$) & ($M_J$)  & ($M_J$) & ($T_{\rm eff}\geq$)
}
\startdata
PG 0132+254   & 430  & $>5$  & $>10$ & $>15$  & $>40$ & 800 K\\
PG 0237+242   & 2110 & $>50$ & $>90$ & $>100$ & $>100$ & \nodata \\
WD 0339+523   & 250  & $>5$  & $>10$ & $>15$ & $>40$ & 800 K\\
WD 0437+152   & 370  & $>6$ & $>10$ & $>20$ & $>40$ & 800 K\\
WD 0453+418   & 110  & $>4$ & $>9$ & $>12$ & $>30$ & 700 K\\
WD 0509$-$007 & 280  & $>10$ & $>40$ & $>50$ & $>72$ & 1600 K\\
PG 0808+595   & 570  & $>9$ & $>20$ & $>30$ & $>60$ & 1100 K\\
PG 0943+441   & 70   & $>2$ & $>7$ & $>10$ & $>20$ & 700 K\\
PG 1229$-$013 & 140  & $>2$ & $>6$ & $>9$ & $>20$ & 700 K\\
PG 1654+637   & 240  & $>2$ & $>5$ & $>7$ & $>15$ & 700 K\\
PG 2226+061   & 160  & $>4$ & $>10$ & $>15$ & $>30$ & 700 K\\
\enddata
\tablecomments{The mass limits are based on the models of I. Baraffe (2009, private
communication; Baraffe et al. 2003). The temperature limits for brown dwarf companions
(the last column) are based on the models of \citet{burrows06}.}
\end{deluxetable}

\subsection{HS 1653+7753}

One of our targets, HS 1653+7753, shows excess flux in the mid-infrared. The contour maps of the IRAC images
show that there is a nearby faint source in the IRAC beam. The 2MASS images do not reveal any nearby, faint sources.
We obtained MMIRS $JHK$ imaging of HS 1653+7753
to resolve this nearby source. Figure 4 displays $JHK$ and IRAC 4.5$\mu$m images of the immediate field around
HS 1653+7753. Our MMIRS data shows that there is a faint source to the southeast of the WD at $2.2\arcsec$ separation.
This source is relatively bright in the mid-infrared, and contaminates the IRAC photometry even in a 2 pixel aperture.
We use a 3 pixel ($3.6\arcsec$) aperture to measure the total flux from the two sources in our IRAC data.

\begin{figure}[b]
\plotone{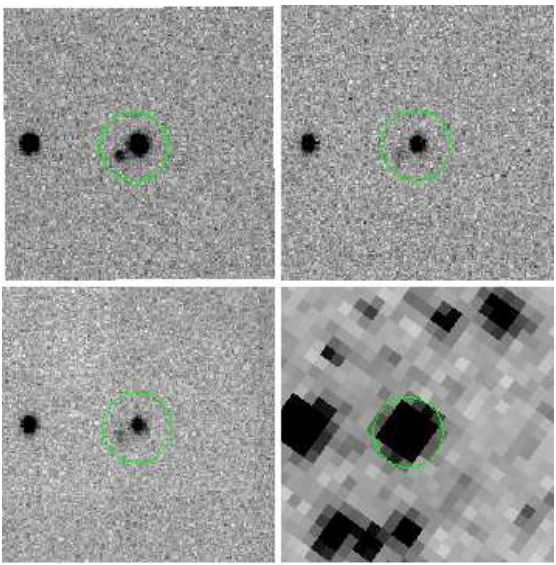}
\caption{MMIRS J, H, K-band and IRAC 4.5 $\mu$m (top left to bottom right) images of the immediate field
around HS 1653+7753. North is up and East is to the left. The image covers $28\arcsec \times 28\arcsec$.
The circle marks a 3.6$\arcsec$ aperture centered on the WD.}
\end{figure}

The nearby source is clearly resolved in the $J$-band data, but it is not detected well in the $H$ and $K$ band images. 
This prevents us from using point spread function fitting photometry.
The FWHM of the near-infrared images is $1.2\arcsec$ (6 pixels). We use a $1.2\arcsec$ aperture to measure the flux from the WD only, and
a $3.6\arcsec$ aperture to measure the total flux from the two sources.
The $6.8\arcmin \times 6.8\arcmin$ field of view of MMIRS enables us to use about 20 nearby 2MASS stars to calibrate the photometry.
We checked the photometry for several 2MASS stars using $1.2\arcsec$ and $3.6\arcsec$ apertures, and found them to be consistent within the errors.
We measure $J=15.80 \pm 0.01$, $H=15.91 \pm 0.02$, and $K=16.06 \pm 0.03$ mag for the WD (using a $1.2\arcsec$ aperture), and
$J=15.70 \pm 0.02$, $H=15.71 \pm 0.05$, and $K=15.83 \pm 0.07$ mag for the total flux in a $3.6\arcsec$ aperture.
The latter are consistent with 2MASS photometry for HS 1653+7753 within the errors.

\begin{figure}
\includegraphics[angle=0,scale=.45]{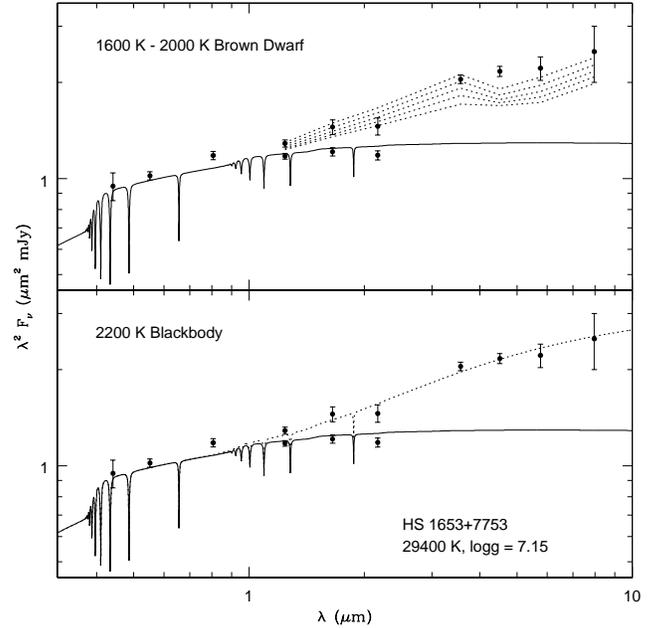}
\caption{Spectral energy distribution of HS 1653+7753.
The near-infrared photometry using both 1.2$\arcsec$ (lower points) and 3.6$\arcsec$ (upper points) apertures is shown.
The solid lines show the expected photospheric flux from the WD. The dashed lines
show the combined flux from the WD and 1600 K - 2000 K brown dwarf companions (top panel),
and a background object with a blackbody temperature of 2200 K (bottom panel).}
\end{figure}

Figure 5 presents the spectral energy distribution of HS 1653+7753. We show the near-infrared photometry for both
$1.2\arcsec$ and $3.6\arcsec$ apertures, and the mid-infrared photometry for a $3.6\arcsec$ aperture.
The combined flux from the WD and possible brown dwarf companions \citep{burrows06} are shown in the top panel.
If the two objects are bound, the observed mid-infrared flux excess implies a relatively warm brown dwarf.
However, the excess flux (at 4.5 $\mu$m) is not explained well by these models.
The bottom panel in Figure 5 presents the combined flux from the WD and a background object
with a blackbody temperature of 2200 K. The mid-infrared flux excess is explained better by a 2200 K background object.
Therefore, the resolved object near the WD is most likely a background source. 

HS 1653+7753 has $\mu_{RA}=-108 \pm 32$ and
$\mu_{DEC}= 118 \pm 99$ mas yr$^{-1}$ in the USNO-B catalog \citep{monet03}. Using 2MASS and MMIRS $J$-band images, we measure
$\mu_{RA}=-19 \pm 15$ and $\mu_{DEC}=76 \pm 19$ mas yr$^{-1}$. If HS 1653+7753 and the nearby source are not bound,
they will be moving further apart. Follow up $J$-band observations in a few years will show if this system is bound.

\subsection{PG 2257+162}

PG 2257+162 is classified as a tentative WD + low-mass MS binary by \citet{wachter03} based on
its 2MASS near-infrared colors.
It shows significant flux excesses in the near- and mid-infrared, and it is not resolved in the IRAC images.
The spectral energy distribution of this WD is presented in Figure 6. It is clear from this figure that the excess comes from
an object with a blackbody temperature of 3300 K. If the unresolved companion is physically associated with the WD, it is most likely
a low-mass stellar companion. We are currently obtaining radial velocity observations of this WD using the 1.5m Tillinghast
telescope to see if this system is a physical binary or not.

\begin{figure}
\plotone{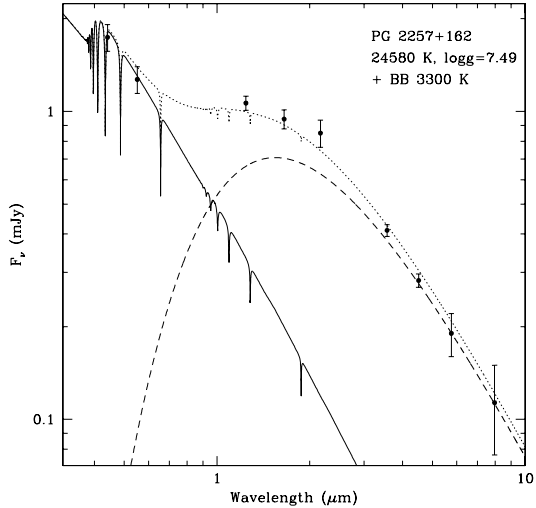}
\caption{Spectral energy distribution of PG 2257+162. The star is unresolved in the IRAC images.
The excess infrared flux is likely due to a late type star companion.
The solid, dashed, and dotted lines show the expected flux for the WD, a 3300 K blackbody, and the combination of the two,
respectively.}
\end{figure}

\section{DISCUSSION AND CONCLUSIONS}

Our IRAC survey of 14 low-mass WDs did not reveal any substellar or planetary mass companions. One of our targets, HS 1653+7753,
has a nearby source that is likely to be a background object. Another target, PG 2257+162, possibly has a low-mass stellar companion.
In addition, the IRAC photometry for one of our targets, WD 0950$-$572 is affected by a nearby star.
The remaining 11 low-mass WDs do not show any mid-infrared flux excesses. Adding the three low-mass WDs observed by
\citet{hansen06} and \citet{mullally07} to our sample brings the sample size to 14 WDs with no detectable mid-infrared flux excess.
Eleven of these targets do not show radial velocity variations, ruling out companions more massive than $\approx0.08M_\odot$
\citep{maxted00}.
Therefore, these 11 low-mass WDs appear to have no stellar (including another WD) or substellar
companions to the limits given in Table 3.
The non-detection of stellar, substellar, or planetary companions to our targets is not inconsistent with the idea
that single low-mass WDs descend from single metal-rich stars through extreme mass loss.

\citet{nelemans98} predict that if a substellar object with $M<32 M_J$ 
goes through a common-envelope phase with a star, it will either evaporate or
spiral-in during the process. More massive objects are expected to survive this
evolution, as evidenced by WD 0137$-$049B \citep[$0.053~M_\odot$,][]{maxted06},
although the theoretical 32 $M_J$ limit needs to be tested by observations.
\citet{nelemans98} use WD 1614+136 to demonstrate that their common envelope phase scenario is feasible. They assume that WD 1614+136 is a
remnant of a 1 $M_{\odot}$ star with a core-mass of 0.33$M_{\odot}$. They find that the minimum companion mass required to expel the envelope
of the progenitor is 21 $M_J$. No companion is detected in IRAC observations of this star, and \citet{hansen06} rule out
$M\geq20M_J$ companions for ages $\leq1$ Gyr. This limit goes up to about 40 $M_J$ for 5 Gyr old companions. 
A well tuned scenario involving a common envelope phase between a $\sim$5 Gyr old 1 $M_{\odot}$ star
and a 21-40 $M_J$ brown dwarf at a certain orbital separation can explain the single, low-mass WD 1614+136. 

Our survey is sensitive to massive brown dwarf companions around the majority of our targets even at 5 Gyr (see Table 3).
We rule out $M>40 M_J$ companions around 8 of our targets at 5 Gyr. Like WD 1614+136, only a well tuned common-envelope phase
scenario involving 20-40 $M_J$ companions would explain the low mass WDs in our sample.
\citet{grether06} find that
the driest part (the minimum number) of the brown dwarf desert occurs at 31 $M_J$. Thus, it seems unlikely that the
scenario proposed by \citet{nelemans98} will explain all single, low-mass WDs. However, given the caveats in understanding
the common-envelope phase evolution and our age dependent limits on possible companions, we cannot rule out this scenario.

An alternative scenario for the formation of single low-mass WDs involves binary mergers of even lower mass WDs \citep{iben97}. \citet{maxted98}
find that such mergers result in rotational velocities on the order of 1000 km s$^{-1}$, unless the WDs are efficient in loosing angular momentum
through mass loss during the merging process. \citet{maxted98} measure rotational velocities of 50 km s$^{-1}$ for two single low-mass WDs and
they suggest that efficient angular momentum loss should take place if these WDs formed through binary mergers. \citet{nelemans98} find this
formation scenario unlikely. Currently, there are no known binary low-mass WD progenitor systems with combined mass $\approx0.4M_{\odot}$ and
with merger time shorter than a Hubble time \citep[see Table 1 in][]{nelemans05}.
Radial velocity studies of extremely low-mass ($0.2M_{\odot}$) WDs are therefore important
to search for such systems \citep{kilic07a,kilic09a} and to see if this scenario is likely
to explain at least a fraction of single, low-mass WDs.

\acknowledgements
Support for this work was provided by NASA through the Spitzer Space Telescope Fellowship Program,
under an award from Caltech. We thank the MMIRS commissioning team for obtaining the observations,
and T. von Hippel for a careful reading of an earlier version of this manuscript.
We also thank an anonymous referee for useful suggestions.

\end{document}